\documentclass[12pt]{article}
\usepackage{apalike}\usepackage{url}\usepackage{hyperref}\usepackage{cite}
\hypersetup{
  colorlinks=true,
  linkcolor=blue,
  filecolor=magenta,      
  urlcolor=cyan,
  pdftitle={Overleaf Example},
  pdfpagemode=FullScreen,
}
\usepackage[utf8]{inputenc}\usepackage{enumerate}\usepackage{amsmath}\usepackage{amssymb}\usepackage{amsthm}\usepackage{graphicx}\usepackage[dvipsnames]{xcolor}\usepackage{tikz}\usepackage{bm}\usepackage{proof}\usepackage{amsfonts}\usepackage{tcolorbox}\usepackage{ascmac}\usepackage{fancybox}
\usepackage[all]{xy}\usepackage{varwidth}\usepackage{etoolbox}\usepackage{siunitx}\usepackage{physics}
\AtBeginDocument{\RenewCommandCopy\qty\SI}  %% told to add this by compiler 7/27/2023
\hypersetup{
    colorlinks,
    citecolor=black,
    filecolor=black,
    linkcolor=black,
    urlcolor=black
}
\usepackage[edges]{forest}
\usepackage{tree}
\newtheorem{definition}{Definition}[section]
\newtheorem{theorem}{Theorem}[section]

\newtheorem{corollary}{Corollary}[section]

\newtheorem{assumption}{Assumption}[section]

\theoremstyle{definition}
\AtEndEnvironment{remark}{\qed}

\AtBeginDocument{\RenewCommandCopy\qty\SI}

\renewcommand{\P}{\mathbb{P}}

\usepackage[margin=1in, includefoot, footskip=0.5in]{geometry}
\usepackage{bbm}

\usepackage{natbib}
\usepackage{comment}
\excludecomment{en-text}

\newcommand{\indep}{\perp \!\!\! \perp}

\newcommand{\E}{{\text{E}}}
\renewcommand{\P}{{\text{P}}}

\begin{document}
\allowdisplaybreaks

\title{\Large The Informativeness of Combined Experimental and\\ Observational Data under Dynamic Selection\thanks{\setlength{\baselineskip}{4mm}We would like to thank Raj Chetty for generously allowing us to use his data sets for our empirical application.\smallskip}}
\author{
Yechan Park\thanks{\setlength{\baselineskip}{4mm}Opportunity Insights, Harvard University, 1280 Massachusetts Avenue, Cambridge, MA 02138 Email: \texttt{yechanpark@fas.harvard.edu}\smallskip}
\and
Yuya Sasaki\thanks{\setlength{\baselineskip}{4mm}Brian and Charlotte Grove Chair and Professor of Economics. Department of Economics, Vanderbilt University, VU Station B \#351819, 2301 Vanderbilt Place, Nashville, TN 37235-1819 Email: \texttt{yuya.sasaki@vanderbilt.edu}}
}
\date{}

\maketitle
\begin{abstract}\setlength{\baselineskip}{6mm}
% This paper studies the problem of identifying the average treatment effect on the treated survivors \citep[ATETS;][]{vikstrom2018bounds} when long-term experimental data are unavailable but instead long-term observational data are available.
% We establish two theoretical results.
% First, it is impossible to obtain informative bounds for the ATETS with no model restriction and no auxiliary data. 
% Second, however, a researcher may enjoy informative bounds if she has access to short-term experimental data in addition.
% To establish the latter, we take advantage of the recent econometric methods to utilize combined experimental and observational data \citep[e.g.,][]{athey2019surrogate,athey2020combining}.
% This result demonstrates that combined experimental and observational data are useful even under dynamic selection.
% The identified set under data combination and a general set of model restrictions are characterized based on \citet{chesher2017generalized}.
% Applying the proposed method, we explore what can be learned about the long-run effects of job training programs on employment without long-term experimental data.

This paper addresses the challenge of estimating the Average Treatment Effect on the Treated Survivors \citep[ATETS;][]{vikstrom2018bounds} in the absence of long-term experimental data, utilizing available long-term observational data instead.
We establish two theoretical results.
First, it is impossible to obtain informative bounds for the ATETS with no model restriction and no auxiliary data. Second, to overturn this negative result, we explore as a promising avenue the recent econometric developments in combining experimental and observational data \citep[e.g.,][]{athey2019surrogate,athey2020combining}; we indeed find that exploiting short-term experimental data can be informative without imposing classical model restrictions.
Furthermore, building on \citet{chesher2017generalized}, we explore how to systematically derive sharp identification bounds, exploiting both the novel data-combination principles and classical model restrictions.
Applying the proposed method, we explore what can be learned about the long-run effects of job training programs on employment without long-term experimental data.
% We apply this methodology to analyze the long-term employment impacts of job training programs without relying on long-term experimental data.


%\medskip\noindent
%{\bf JEL Codes:}
\end{abstract}

\newpage

%%%%%%%%%%%%%%%%%%%%%%%%%%%%%%%%%%%%%%%
\section{Introduction}
%%%%%%%%%%%%%%%%%%%%%%%%%%%%%%%%%%%%%%%
In assessing policy impacts, precise estimation of long-term effects is crucial. 
For example, when a policy concerns active labor market policies, one of the most frequently investigated questions is how the policy influences the transition from unemployment to employment in the long run; there is an extensive list of studies that have explored duration models of unemployment following active labor market policies \citep[e.g.,][]{ham1996effect,abbring2003nonparametric,gritz1993impact,card1987measuring,ridder1986event,van2001active}.

A na\"ive model for estimating long-term effects might assume direct comparability between treated and untreated groups over time. However, such an approach overlooks dynamic selection, where the initial comparability assumed in experimental settings with random treatment assignment, will not generally extend into the future due to changes in the composition of the groups. 
Dynamic selection challenges the identifiability of treatment effects as such. 

A notable innovation was made by \cite{vikstrom2018bounds} who introduced the Average Treatment Effect on the Treated Survivors (ATETS) to address this point.
It aims to estimate long-term effects by maintaining comparability for potential outcomes.
Yet, there is a non-trivial data requirement here.
\cite{vikstrom2018bounds} assume that there are long-term observed randomized experiments, and explore partial identification of the ATETS. 
Practically, collecting experimental data is both costly and time-consuming. 
This drawback is exacerbated when a researcher is interested in longer-term outcomes, and it is sometimes even infeasible to collect such data due to loss of follow-up. 

In light of such a limitation that long-term experimental data are hardly available in practice, the recent econometrics literature proposes to combine short-term experimental data with long-term observational data \citep[e.g.,][]{athey2019surrogate,athey2020combining,imbens2022long,ghassami2022combining}. 
This novel approach paves the way for a wider spectrum of empirical research as long-term observational data are much more widely available than long-term experimental data.
We exploit this novel approach to address the above problem of partially identifying the ATETS without relying on long-term experiments.

While the aforementioned existing papers \citep{athey2019surrogate,athey2020combining,imbens2022long,ghassami2022combining} on data combination focus on the average treatment effects (ATE) and those on the treated (ATT), their point-identification results do not extend to the ATETS, which measures the average treatment effect focusing on a subpopulation of disadvantaged individuals of policy makers' interest.
Even in the presence of long-term experimental data, the ATETS is only partially identified in general \citep[cf.][]{vikstrom2018bounds}.
In the absence of long-term experimental data, therefore, it is not surprising that the ATETS is only partially identified.\footnote{Indeed, there exists an assumption which enables point identification -- see Appendix \ref{sec:point_identification} for formal results. However, such an assumption is deemed to be implausible in the literature on employment dynamics among others, and hence we abstract from its exploration in the main part of this paper.}
Yet, we show that data combination yields informative bounds.

We are not the first to investigate partial identification under data combination.
First, \citet{fan2014identifying} apply the Fr\'echet-Hoeffding bounds to achieve sharp bounds for the average treatment effect by combining data sets with separate outcome and conditioning covariates.
Our approach incorporates additional complexities associated with observational selection biases and utilizes the complementary quality of experimental versus observational data.
Second, \citet{kallus2022assessing} extend the application of the Fr\'echet-Hoeffding bounds to bound fairness metrics by stitching together two large-scale observational datasets.
Our methodology goes beyond by focusing on long-term policy evaluation and addressing its specific challenges related to dynamic selection.

More generally, we extend beyond the conventional application of Fr\'echet-Hoeffding bounds to that based on \citet{chesher2017generalized}. Our paper thus enriches the data combination discourse, offering enhanced tools for policymakers interested in understanding the long-term effects of interventions on particularly disadvantaged subpopulations under a general set of model restrictions.
Furthermore, to the best of our knowledge, this paper is the first to investigate partial identification under the novel setting of combined experimental and observational data introduced by \citet{athey2019surrogate,athey2020combining} and others.

\section{The Baseline Setup}
%%%%%%%%%%%%%%%%%%%%%%%%%%%%%%%%%%%%%%%

Having observed pre-treatment covariates $X_i$, an experimenter randomly assigns a binary treatment $W_i$ (e.g., job training) to the $i$-th individual, and observes the short-term outcome $Y_{1i}$ (e.g., employment status) of this individual $i$.
A sample for such an experiment is indicated by $G_i=E$.

On the other hand, suppose that an econometrician has access to long-run observational data $\{(Y_{2i}, Y_{1i}, W_i, X_i)\}$, where $Y_{2i}$ is a long-term outcome.
A sample in this observational data is indicated by $G_i=O$.
The following table summarizes the observability of each of the four variables depending on whether $G_i = E$ or $O$.

%%%%%%%%%%%%%%%%%%%%%%%%%%%%%%%%%%%%%%%
    \begin{table}[h]\centering
        \renewcommand{\arraystretch}{0.9}
        \begin{tabular}{c|cccc}
            &$Y_{2i}$&$Y_{1i}$&$X_i$&$W_i$\\\hline    $G_i=E$&$\times$&$\bigcirc$&$\bigcirc$&$\bigcirc$\\\hline
            $G_i=O$&$\bigcirc$&$\bigcirc$&$\bigcirc$&$\bigcirc$
        \end{tabular}
    \end{table}
%%%%%%%%%%%%%%%%%%%%%%%%%%%%%%%%%%%%%%%

\noindent 
This setting inherits from the recent literature on combined experimental and observational data \citep[e.g.,][]{athey2019surrogate,athey2020combining}

We adopt the potential-outcome notations as follows.
Let $Y_{ti}(0)$ and $Y_{ti}(1)$ denote the potential outcomes without and with treatment, respectively, for individual $i$ at time $t$.
Consequently, the observed outcome $Y_{ti}$ can be written in terms of these latent variables as $Y_{ti} = (1-W_i)Y_{ti}(0) + W_i Y_{ti}(1)$.
Hereafter, we omit the individual subscript $i$.

In this setting, we are interested in learning about the average treatment on the treated survivors \citep[ATETS;][]{vikstrom2018bounds} in observational data, defined by
\begin{align}\label{eq:atets}
\theta = \E[Y_2(1)|Y_1(1)=0,G=O] - \E[Y_2(0)|Y_1(1)=0,G=O].
\end{align}
The original paper by \citet{vikstrom2018bounds} focuses on the ATETS in experimental data (indicated by $G=E$), but this is infeasible in our framework in which the long-term outcome $Y_2$ for the experimental units (indicated by $G=E$) is missing.
Furthermore, since the experimental random assignment is not guaranteed for the observational units (indicated by $G=O$) in our framework, the existing partial identification strategy of \citet{vikstrom2018bounds} will not apply to \eqref{eq:atets}.

%%%%%%%%%%%%%%%%%%%%%%%%%%%%%%%%%%%%%%%
\section{No Restriction And No Auxiliary Data}\label{sec:no_no}
%%%%%%%%%%%%%%%%%%%%%%%%%%%%%%%%%%%%%%%

This section presents what we can learn about the ATETS, $\theta$, when a researcher has access to only the long-term observational data $(Y_2,Y_1,X,W)|G=O$ and no restriction is imposed on the underlying model.
It turns out that the result is negative as formally stated below.

%%%%%%%%%%%%%%%%%%%%%%%%%%%%%%%%%%%%%%%
\begin{theorem}\label{theorem:non_informative}
Suppose that there is no model restriction and a researcher has access to only observational data indicated by $G=O$.
\\
(i) The sharp bounds for $\E[Y_2(1)|Y_1(1)=0,G=O]$ are given by 
\begin{align*}
\frac{\P(Y_2=1,Y_1=0,W=1|G=O)}{\P(Y_1=0,W=1|G=O)+\P(W=0|G=O)}
\leq
\E[Y_2(1)|Y_1(1)=0,G=O]&
\\
\leq
\frac{\P(Y_2=1,Y_1=0,W=1|G=O)+\P(W=0|G=O)}{\P(Y_1=0,W=1|G=O)+\P(W=0|G=O)}&.
\end{align*}
(ii) The sharp bounds for $\E[Y_2(0)|Y_1(1)=0,G=O]$ are given by $$0 \leq \E[Y_2(0)|Y_1(1)=0,G=O] \leq 1.$$
\end{theorem}
%%%%%%%%%%%%%%%%%%%%%%%%%%%%%%%%%%%%%%%

Our proof of this theorem is based on an application of \citet[][Theorem 1]{horowitz2000nonparametric}, and can be found in Appendix \ref{sec:theorem:non_informative}.
Part (i) (respectively, part (ii)) provides sharp bounds for the first term (respectively, second term) on the right-hand side of \eqref{eq:atets}.
Part (ii) implies that the sharp bounds for the second term are not informative.
Consequently, this theorem implies that, with no model restriction and no auxiliary data, there are no informative bounds for $\theta$ in the sense of \citet{horowitz2000nonparametric} and \citet{manski2003partial}.

%%%%%%%%%%%%%%%%%%%%%%%%%%%%%%%%%%%%%%%
\begin{definition}\label{def:informative}
Bounds $\theta^L \leq \theta \leq \theta^U$ are informative if $\theta^U - \theta^L < 1$.
\end{definition}
%%%%%%%%%%%%%%%%%%%%%%%%%%%%%%%%%%%%%%%

%%%%%%%%%%%%%%%%%%%%%%%%%%%%%%%%%%%%%%%
\begin{corollary}\label{corollary:no_informative}
Suppose that there is no model restriction and a researcher has access to only observational data indicated by $G=O$.
Then there are no informative bounds for $\theta$.
\end{corollary}
%%%%%%%%%%%%%%%%%%%%%%%%%%%%%%%%%%%%%%%

%%%%%%%%%%%%%%%%%%%%%%%%%%%%%%%%%%%%%%%
\section{Combined Experimental And Observational Data}\label{sec:combined}
%%%%%%%%%%%%%%%%%%%%%%%%%%%%%%%%%%%%%%%

This section presents what we can learn about the ATETS, $\theta$, when a researcher has access to short-run experimental data $(Y_1,X,W)|G=E$ in addition to the
 the long-term observational data $(Y_2,Y_1,X,W)|G=O$.

To make use of combined experimental and observational data, researchers often invoke the following assumptions of internal and external validity, e.g., \citet{athey2019surrogate} and \citet{athey2020combining}.
%%%%%%%%%%%%%%%%%%%%%%%%%%%%%%%%%%%%%%%
\begin{assumption}[Experimental Internal Validity]\label{ass:IV}
    ${}$\\${}$\hspace{1cm}
    $W \indep Y_2(w), Y_1(w) | X, G=E$ for each $w \in \{0,1\}$.
\end{assumption}
%%%%%%%%%%%%%%%%%%%%%%%%%%%%%%%%%%%%%%%

%%%%%%%%%%%%%%%%%%%%%%%%%%%%%%%%%%%%%%%
\begin{assumption}[External Validity of the Experiment]\label{ass:EV}
    ${}$\\${}$\hspace{1cm}
    $G \indep Y_2(w), Y_1(w) | X$ for each $w \in \{0,1\}$.
\end{assumption}
%%%%%%%%%%%%%%%%%%%%%%%%%%%%%%%%%%%%%%%

\noindent
The following assumption of latent unconfoundedness (LU) is also proposed by the existing literature on data combination.

%%%%%%%%%%%%%%%%%%%%%%%%%%%%%%%%%%%%%%%
\begin{assumption}[Latent Unconfoundedness; \citealp{athey2020combining}] \label{ass:LU}
    ${}$\\${}$\hspace{1cm}
    $W \indep Y_2(w) \mid Y_1(w), X, G=O$ for each $w \in \{0,1\}$.
\end{assumption}
%%%%%%%%%%%%%%%%%%%%%%%%%%%%%%%%%%%%%%%

In addition to these independence conditions, we assume the strong overlap condition that the propensity score $(x,g) \mapsto \P(W=1|X=x,G=g)$ is bounded away from 0 and 1 throughout.

Recall that we have no informative bounds for the second term $\E[Y_2(0)|Y_1(1)=0,G=O]$ of the ATETS on the right-hand side of \eqref{eq:atets} under no restriction and no auxiliary data -- see Theorem \ref{theorem:non_informative} (ii) and Corollary \ref{corollary:no_informative}.
The following theorem, on the other hand, shows that data combination along with the commonly invoked assumptions in this literature allows for possibly informative bounds of it.

%%%%%%%%%%%%%%%%%%%%%%%%%%%%%%%%%%%%%%%
\begin{theorem}\label{theorem:data_combination2}
If Assumptions \ref{ass:IV}, \ref{ass:EV}, and \ref{ass:LU} are satisfied, then the sharp bounds of $\E[Y_2(0)|Y_1(1)=0,G=O]$ are given by\small
\begin{align*}
\max\left\{\frac{\E[\E[\E[Y_2|Y_1,W=0,X,G=O]|X,G=E]|G=O] - \E[\E[Y_1|W=1,X,G=E]|G=O]}{1-\E[\E[Y_1|W=1,X,G=E]|G=O]},0\right\}&
\\
\leq
\E[Y_2(0)|Y_1(1)=0,G=O]
\leq
\min\left\{\frac{\E[\E[\E[Y_2|Y_1,W=0,X,G=O]|X,G=E]|G=O]}{1-\E[\E[Y_1|W=1,X,G=E]|G=O]},1\right\}&.
\end{align*}\normalsize
\end{theorem}
%%%%%%%%%%%%%%%%%%%%%%%%%%%%%%%%%%%%%%%

Our proof of this theorem is based on combined applications of Fr\'echet-Hoeffding Theorem and \citet{athey2020combining}, and can be found in Appendix \ref{sec:theorem:data_combination2}.
We apply Fr\'echet-Hoeffding Theorem on the conditional copula between $Y_2(0)$ and $Y_1(1)$ given $G=O$.
Note that Assumptions \ref{ass:IV}, \ref{ass:EV}, and \ref{ass:LU} do \textit{not} impose any restriction on this conditional copula between $Y_2(0)$ and $Y_1(1)$ given $G=O$.
Therefore, the sharpness of the Fr\'echet-Hoeffding bounds persists even after imposing these three assumptions.

The below corollary formalizes the necessary and sufficient condition for informative bounds.
%%%%%%%%%%%%%%%%%%%%%%%%%%%%%%%%%%%%%%%
\begin{corollary}\label{corollary:informative}
If Assumptions \ref{ass:IV}, \ref{ass:EV}, and \ref{ass:LU} are satisfied, then the sharp bounds for $\E[Y_2(0)|Y_1(1)=0,G=O]$ are informative if and only if
\begin{align*}
    \E[Y_2(0)|G=O] > \E[Y_1(1)|G=O] \quad\text{ or }\quad \E[Y_2(0)|G=O] + \E[Y_1(1)|G=O] < 1.
\end{align*}
\end{corollary}
%%%%%%%%%%%%%%%%%%%%%%%%%%%%%%%%%%%%%%%

Our proof of this corollary can be found in Appendix \ref{sec:corollary:informative}, and follows immediately from our proof of Theorem \ref{theorem:data_combination2} which is presented in Appendix \ref{sec:theorem:data_combination2}.
Since this necessary and sufficient condition can be satisfied in practice, there is a chance that the data combination may yield informative bounds, unlike the case in which a researcher has access to only observational data -- see Section \ref{sec:no_no}.

We remark that 
$\E[Y_1(1)|G=O]$ and $\E[Y_2(0)|G=O]$ showing up in the necessary and sufficient condition in Corollary \ref{corollary:informative} are identifiable. 
Specifically, they are identified by \eqref{eq:sharp_bounds:Y11} and \eqref{eq:sharp_bounds:Y20} in the proof of Theorem \ref{theorem:data_combination2} in Appendix \ref{sec:theorem:data_combination2}.
For the data set we use in our empirical application (Section \ref{sec:application_job_training}), this necessary and sufficient condition for the informativeness of the sharp bounds is satisfied.

%%%%%%%%%%%%%%%%%%%%%%%%%%%%%%%%%%%%%%%
\section{General Restrictions}
%%%%%%%%%%%%%%%%%%%%%%%%%%%%%%%%%%%%%%%

The previous section demonstrates that data combination is a practical approach to overturn the negative results in our previous section that solely relies on observational data (cf. Corollary \ref{corollary:no_informative} vs Corollary \ref{corollary:informative}).
%However, the assumptions stated in Section \ref{sec:combined} fail to impose restrictions on the conditional joint distribution $(Y_2(0),Y_1(0))|G=O$, which would be useful to further improve the bounds for the second term, $\E[Y_2(0)|Y_1(1)=0,G=0]$, on the right-hand side of \eqref{eq:atets}.

Since the potential outcomes, $Y_2(0)$ and $Y_1(1)$, concern distinct treatment status, some model restrictions like monotone treatment response \citep[MTR;][]{manski1997monotone} may be useful to improve the bounds if such a restriction is reasonable under an application of interest.
%Likewise, since the potential outcomes, $Y_2(0)$ and $Y_1(1)$, concern distinct time periods, some model restrictions like stationarity may be also useful to improve the bounds if such a restriction is reasonable under an application of interest.
This section introduces a general framework in which such a restriction may be imposed in the context of data combination introduced in Section \ref{sec:combined}.

Let $\mathcal{U} = \{0,1\}^4$ denote the set of values that the vector $U := (Y_2(1),Y_2(0),Y_1(1),Y_1(0))$ of the potential outcomes could possibly take.
Similarly, let $\mathcal{W} = \{0,1\}$, $\mathcal{X}$, and $\mathcal{G}=\{E,O\}$ denote the sets of values that $W$, $X$, and $G$, could possibly take, respectively.
Define $\mathcal{P}$ as the set of functions containing the conditional distribution function $P_{U|WXG}: \mathcal{B}(\mathcal{U}) \times \mathcal{W} \times \mathcal{X} \times \mathcal{G} \rightarrow [0,1]$, where $\mathcal{B}(\mathcal{U}) = 2^{\mathcal{U}}$ denotes the power set of $\mathcal{U}$ (which is also a finite sigma-algebra).

With the short-hand notations
\begin{align*}
\mathcal{U}(Y_2,Y_1,W,E)&=\left\{(y_2(1),y_2(0),y_1(1),y_1(0)) \in \mathcal{U} \left\vert \begin{array}{l}Y_1=(1-W)y_1(0) + Wy_1(1)\end{array} \right.\right\}
\text{ and}\\
\mathcal{U}(Y_2,Y_1,W,O)&=\left\{(y_2(1),y_2(0),y_1(1),y_1(0)) \in \mathcal{U} \left\vert \begin{array}{l}Y_2=(1-W)y_2(0) + Wy_2(1) \\ Y_1=(1-W)y_1(0) + Wy_1(1)\end{array} \right.\right\},
\end{align*}
we define the containment functional $C$ by
\begin{align*}
C(S|w,x,g) = \P(\mathcal{U}(Y_2,Y_1,W,G) \subset S|W=w,X=x,G=g)
\end{align*}
for each $S \in \mathcal{B}(\mathcal{U})$ and $(w,x,g) \in \mathcal{W} \times \mathcal{X} \times \mathcal{G}$.
We can characterize the sharp identified set for $P_{U|WXG}$ as follows.

%%%%%%%%%%%%%%%%%%%%%%%%%%%%%%%%%%%%%%%
\begin{theorem}\label{theorem:general}
The sharp identified set for $P_{U|WXG}$ is given by $\mathcal{P}^\ast =$
\begin{align*}
\left\{Q_{U|WXG} \in \mathcal{P} \left\vert C(S|w,x,g) \leq Q_{U|WXG}(S|w,x,g) \text{ for all } S \in \mathcal{B}(\mathcal{U}), (w,x,g) \in \mathcal{W} \times \mathcal{X} \times\mathcal{G} \right.\right\}.
\end{align*}
\end{theorem}
%%%%%%%%%%%%%%%%%%%%%%%%%%%%%%%%%%%%%%%

\noindent
This theorem follows from a direct application of Corollary 1 of \citet{chesher2017generalized}.
At first glance, Theorem \ref{theorem:general} appears to be silent about ways to impose assumptions or restrictions, such as Assumptions \ref{ass:IV}--\ref{ass:LU} or the MTR.
However, we can naturally embed such restrictions in the definition of $\mathcal{P}$.
Sections \ref{sec:imposing_IV_EV} and \ref{sec:MTR} concretely propose those steps.

Once we obtain the sharp identified set $\mathcal{P}^\ast$ for $P_{U|WXG}$ characterized by Theorem \ref{theorem:general}, one can in turn obtain the sharp identified set for the ATETS, $\theta$, as follows. 

%%%%%%%%%%%%%%%%%%%%%%%%%%%%%%%%%%%%%%%
\begin{corollary}\label{corollary:identified_set}
The sharp identified set for $\theta$ is given by\small
\begin{align*}
\Biggl\{\left. 
\frac{\sum_{y_2(1),y_2(0),y_1(0),w,x} (y_2(1)-y_2(0)) Q_{U|WXG}(\{(y_2(1), y_2(0),0, y_1(0))\}|w,x,O)) f_{WX|G}(w,x|O)}{\sum_{y_2(1),y_2(0),y_1(0),w,x} Q_{U|WXG}(\{( y_2(1), y_2(0),0, y_1(0))\}|w,x,O))f_{WX|G}(w,x|O)}
\right\vert&
\\
Q_{U|WXG} \in \mathcal{P}^\ast
&\Biggr\},
\end{align*}\normalsize
where $f_{WX|G}$ denotes the conditional probability mass function.
\end{corollary}
%%%%%%%%%%%%%%%%%%%%%%%%%%%%%%%%%%%%%%%

\noindent
Appendix \ref{sec:computation} presents practical details on how to compute the bounds.

%%%%%%%%%%%%%%%%%%%%%%%%%%%%%%%%%%%%%%%
\subsection{Imposing the Restrictions on Data Combination}\label{sec:imposing_IV_EV}
%%%%%%%%%%%%%%%%%%%%%%%%%%%%%%%%%%%%%%%

This section demonstrates how to restrict $\mathcal{P}$ by incorporating the assumptions for data combination (Assumptions \ref{ass:IV}--\ref{ass:LU}).

We can impose experimental internal validity (Assumption \ref{ass:IV}) by restricting $\mathcal{P}$ with the requirement that
\begin{align}\label{eq:IV}
Q_{U|WXG}(\{u\}|0,x,E) = Q_{U|WXG}(\{u\}|1,x,E).
\end{align}
for all $u \in \mathcal{U}$, $x \in \mathcal{X}$, and $Q_{U|WXG} \in \mathcal{P}$.

Similarly, we can impose external validity (Assumption \ref{ass:EV}) by restricting $\mathcal{P}$ with the requirement that
\begin{align}\label{eq:EV}
\sum_{w=0}^1 Q_{U|WXG}(\{u\}|w,x,E)f_{W|XG}(w|x,E) = \sum_{w=0}^1 Q_{U|WXG}(\{u\}|w,x,O) f_{W|XG}(w|x,O).
\end{align}
for all $u \in \mathcal{U}$, $x \in \mathcal{X}$, and $Q_{U|WXG} \in \mathcal{P}$, where $f_{W|XG}$ denotes the conditional probability mass function.

Finally, we can impose the latent unconfoundedness \citep[LU;][]{athey2020combining} condition (Assumption \ref{ass:LU}) by restricting $\mathcal{P}$ with the requirement that
\begin{align}
&\frac{\sum_{\widetilde y_2(1)=0}^1 \sum_{\widetilde y_1(1)=0}^1 Q_{U|WXG}(\{(\widetilde y_2(1),y_2(0),\widetilde y_1(1),y_1(0))\}|1,x,O)}{\sum_{\widetilde y_2(1)=0}^1 \sum_{\widetilde y_2(0)=0}^1 \sum_{\widetilde y_1(1)=0}^1 Q_{U|WXG}(\{(\widetilde y_2(1),\widetilde y_2(0),\widetilde y_1(1),y_1(0))\}|1,x,O)}
\notag\\
&=
\frac{\sum_{\widetilde y_2(1)=0}^1 \sum_{\widetilde y_1(1)=0}^1 Q_{U|WXG}(\{(\widetilde y_2(1),y_2(0),\widetilde y_1(1),y_1(0))\}|0,x,O)}{\sum_{\widetilde y_2(1)=0}^1 \sum_{\widetilde y_2(0)=0}^1 \sum_{\widetilde y_1(1)=0}^1 Q_{U|WXG}(\{(\widetilde y_2(1),\widetilde y_2(0),\widetilde y_1(1),y_1(0))\}|0,x,O)}
\label{eq:lu1}
\end{align}
for all $(y_2(0),y_1(0)) \in \{0,1\}^2$, $x \in \mathcal{X}$, and $Q_{U|WXG} \in \mathcal{P}$, and
\begin{align}
&\frac{\sum_{\widetilde y_2(0)=0}^1 \sum_{\widetilde y_1(0)=0}^1 Q_{U|WXG}(\{(y_2(1),\widetilde y_2(0),y_1(1),\widetilde y_1(0))\}|1,x,O)}{\sum_{\widetilde y_2(1)=0}^1 \sum_{\widetilde y_2(0)=0}^1 \sum_{\widetilde y_1(0)=0}^1 Q_{U|WXG}(\{ (\widetilde y_2(1),\widetilde y_2(0),y_1(1),\widetilde y_1(0)) \}|1,x,O)}
\notag\\
&=
\frac{\sum_{\widetilde y_2(0)=0}^1 \sum_{\widetilde y_1(0)=0}^1 Q_{U|WXG}(\{(y_2(1),\widetilde y_2(0),y_1(1),\widetilde y_1(0))\}|0,x,O)}{\sum_{\widetilde y_2(1)=0}^1 \sum_{\widetilde y_2(0)=0}^1 \sum_{\widetilde y_1(0)=0}^1 Q_{U|WXG}(\{(\widetilde y_2(1),\widetilde y_2(0),y_1(1),\widetilde y_1(0))\}|0,x,O)}
\label{eq:lu2}
\end{align}
for all $(y_2(1),y_1(1)) \in \{0,1\}^2$, $x \in \mathcal{X}$, and $Q_{U|WXG} \in \mathcal{P}$.

%%%%%%%%%%%%%%%%%%%%%%%%%%%%%%%%%%%%%%%
%\subsection{Imposing Model Restrictions}
%%%%%%%%%%%%%%%%%%%%%%%%%%%%%%%%%%%%%%%

%%%%%%%%%%%%%%%%%%%%%%%%%%%%%%%%%%%%%%%
\subsection{Imposing Model Restrictions}\label{sec:MTR}
%%%%%%%%%%%%%%%%%%%%%%%%%%%%%%%%%%%%%%%

In addition to the restrictions on data combination as presented in the previous subsection, we may impose common model restrictions used in the econometrics literature.
While there are many important examples, we feature the monotone treatment response \citep[MTR;][]{manski1997monotone} in this section due to its wide use in the partial identification literature as well as its relevance to many applications.

Define the MTR in the context of our model as follows:
\begin{align}\label{eq:MTR}
Y_t(1) \ge Y_t(0)    
\end{align}
almost surely given $(W,X,G)=(w,x,g)$ for each $t \in \{1,2\}$ and $(w,x,g) \in \mathcal{W} \times \mathcal{X} \times \mathcal{G}$.

We can impose this MTR condition \eqref{eq:MTR} by restricting $\mathcal{P}$ with the requirement that 
\begin{align}\label{eq:MTR1}
Q_{U|WXG}(\{(y_2(1),y_2(0),0,1) \} |w,x,g) = 0
\end{align}
for all $(y_2(1),y_2(0)) \in \{0,1\}^2$, $w \in \mathcal{W}$, $x \in \mathcal{X}$, $g \in \mathcal{G}$, and $Q_{U|WXG} \in \mathcal{P}$, and
\begin{align}\label{eq:MTR2}
Q_{U|WXG}(\{(0,1,y_1(1),y_1(0)) \}|w,x,g) = 0
\end{align}
for all $(y_1(1),y_1(0)) \in \{0,1\}^2$, $w \in \mathcal{W}$, $x \in \mathcal{X}$, $g \in \mathcal{G}$, and $Q_{U|WXG} \in \mathcal{P}$.

While we focus on the MTR in this section for an illustration, we emphasize again that other model restrictions may be imposed in a similar manner -- see Appendix \ref{sec:additional_model_selections} for three additional examples.

\section{Empirical Application}\label{sec:application_job_training}
%%%%%%%%%%%%%%%%%%%%%%%%%%%%%%%%%%%%%%%

This section demonstrates the bounds using real data consisting of combined experimental and observational data.
We analyze the long-run treatment effects of job training programs on employment.
The same data sets as \citet{athey2019surrogate} are used to this end but we provide a brief description of them for the convenience of readers in Appendix \ref{sec:data_job_training}.

%%%%%%%%%%%%%%%%%%%%%%%%%%%%%%%%%%%%%%%
\subsection{Key Variables}
%%%%%%%%%%%%%%%%%%%%%%%%%%%%%%%%%%%%%%%

We focus on the following variables.
First, the long-term outcome \textbf{$Y_2$} represents a binary employment status in the long term, defined by employment in the ninth year (one of the 33rd through the 36th quarters).
Second, the short-term outcome \textbf{$Y_1$} represents a binary employment status in the short term, defined by employment in the first year (one of the 1st through the 4th quarters).
Third, \textbf{$W$} is a binary indicator of treatment in the form of an assignment to the job training program.
The experimental units are indicated by $G=E$ while observational units are indicated by $G=O$.

%%%%%%%%%%%%%%%%%%%%%%%%%%%%%%%%%%%%%%%
\subsection{Results}
%%%%%%%%%%%%%%%%%%%%%%%%%%%%%%%%%%%%%%%

We compute estimates of the bounds of $\theta$ under various combinations of restrictions, including the internal validity (IV; Assumption \ref{ass:IV}), external validity (EV; Assumption \ref{ass:EV}), monotone treatment response (MTR; Section \ref{sec:MTR}), latent unconfoundedness (LU; Assumption \ref{ass:LU}), and no state dependence (NSD; Assumption \ref{ass:NSD} -- see Appendix \ref{sec:point_identification}).
As argued in Appendix \ref{sec:point_identification}, we treat the NSD condition as an implausible example, unlike the other restrictions.
The first three (IV, EV, and LU) are restrictions on data combination, whereas the last two (MTR and NSD) are model restrictions.
Table \ref{tab:results} summarizes the results.

\begin{table}[t]
\centering
\begin{tabular}{rcccccccrl}
\hline\hline
& \multicolumn{3}{c}{Data} && \multicolumn{2}{c}{Model} && \multicolumn{2}{c}{}\\
& \multicolumn{3}{c}{Combination} && \multicolumn{2}{c}{Restriction} && \multicolumn{2}{c}{Set}\\
\cline{2-4}\cline{6-7}
& IV & EV & LU && MTR & NSD && \multicolumn{2}{c}{Estimate}\\
\hline
(1) & No  & No  & No  && No  & No  && $[-0.77,$ & $0.52]$\\
(2) & Yes & Yes & No  && No  & No  && $[-0.73,$ & $0.43]$\\
(3) & Yes & Yes & Yes && No  & No  && $[-0.53,$ & $0.36]$\\
(4) & No  & No  & No  && Yes & No  && $[0.01,$  & $0.51]$\\
(5) & Yes & Yes & No  && Yes & No  && $[0.01,$  & $0.43]$\\
(6) & Yes & Yes & Yes && Yes & No  && $[0.01,$  & $0.08]$\\
\hline
(7) & Yes & Yes & Yes && No  & Yes && \multicolumn{2}{c}{$\{-0.02\}$}\\
%(8) & Yes & Yes & Yes && Yes & Yes && $\emptyset$\\
\hline\hline
\end{tabular}
\caption{Set estimates of the ATETS of the job training program on employment under various combinations of the restrictions including the internal validity (IV), external validity (EV), monotone treatment response (MTR), latent unconfoundedness (LU), and no state dependence (NSD).}${}$
\label{tab:results}
\end{table}

Row (1) shows the sharp bounds under no restriction and no auxiliary data, as in Section \ref{sec:no_no}.
As implied by Theorem \ref{theorem:non_informative} and Corollary \ref{corollary:no_informative}, these sharp bounds $[-0.77,0.53]$ are not informative in the sense that the length is larger than one.
In particular, it fails to sign the estimate.

Row (2) demonstrates that the data combination contributes to shrinking the bounds from down to $[-0.74,0.43]$. 
Both the lower bound increases and the upper bound decreases.
Row (3) shows that additionally imposing the LU can further shrink the bounds down to $[-0.56,0.37]$.

Next, let us disregard the data combination for the moment.
Row (4) shows that solely imposing the model restriction of MTR significantly improves the lower bound as it effectively rules out negative treatment effects.
However, it does not change the upper bound that much compared to row (1).
To lower the upper bound, the data combination will be useful as we observed above.
Indeed, row (5) shows that using the data combination contributes to decreasing the upper bound.
Furthermore, row (6) demonstrates that imposing the LU as well as MTR under data combination can significantly shrink the bounds.

Finally, row (7) shows that the point estimate based on the NSD restriction, as in Appendix \ref{sec:point_identification}, is $-0.02$.
% This estimate has the least uncertainty, but it should be interpreted as implausibly reducing uncertainty by far the most informative, as a point estimate as opposed to a set estimate, but its negative sign exhibits a qualitative difference from the results under the MTR.
The resulting point exhibits the lowest level of uncertainty. However, its interpretation warrants caution due to its implausibly significant reduction in uncertainty. In particular, the negative sign represents a qualitative departure from the results obtained under the MTR method. 
Indeed, as argued in Section \ref{sec:point_identification}, the NSD condition is considered to be implausible in the existing literature.
\section{Summary}
%%%%%%%%%%%%%%%%%%%%%%%%%%%%%%%%%%%%%%%

This paper studies the problem of partially identifying the average treatment effect on the treated survivors \citep[ATETS;][]{vikstrom2018bounds} when a researcher has no access to long-term experimental data.
Two theoretical results are established.
First, it is impossible to obtain informative bounds for the ATETS if no model restrictions are imposed and no auxiliary data are available.
Second, one may enjoy informative bounds if the researcher has access to long-term observational data to complement short-term experimental data.
To establish the latter result, we take advantage of the recent developments in combined experimental and observational data \citep[e.g.,][]{athey2019surrogate,athey2020combining}.
We characterize the identified set under data combination and a general set of model restrictions based on \citet{chesher2017generalized}.
Applying the proposed method, we explore what can be learned about the long-run effects of job training programs on employment without long-term experimental data, and we discover informative bounds can be obtained under reasonable assumptions. 

This paper shows that combining experimental and observational data is a promising approach to tackling the challenge of static and dynamic selection. Some interesting future work would be to explore whether similar approaches can apply to other traditional econometric challenges such as noncompliance or measurement error.

\appendix
\section*{Appendix}
\section{Mathematical Proofs}
%%%%%%%%%%%%%%%%%%%%%%%%%%%%%%%%%%%%%%%

%%%%%%%%%%%%%%%%%%%%%%%%%%%%%%%%%%%%%%%
\subsection{Proof of Theorem \ref{theorem:non_informative}}\label{sec:theorem:non_informative}
%%%%%%%%%%%%%%%%%%%%%%%%%%%%%%%%%%%%%%%
\begin{proof}
{Part (i):}
Let $Z_1$ (respectively, $Z_2$) indicate the event that a researcher observes $Y_1(1)$ (respectively, $Y_2(1)$).
By the law of total probability and then applying the Bayes' theorem, we have
\begin{align*}
&\E[Y_2(1)|Y_1(1)=0,G=O]=
\sum_{j,k} E_{jk} G_{jk}
=
\frac{\sum_{j,k} E_{jk} Q_{jk} p_{jk}}{\sum_{j,k} Q_{jk} p_{jk}}
\end{align*}
where
$
E_{jk} = \P(Y_2(1)=1|Y_1(1)=0,Z_1=j,Z_2=k,G=O),
$
$
G_{jk} = \P(Z_1=j,Z_2=k|Y_1(1)=0,G=O),
$
$
Q_{jk} = \P(Y_1(1)=0|Z_1=j,Z_2=k,G=O),
$
and
$
p_{jk} = \P(Z_1=j,Z_2=k|G=O).
$
Note that we do not observe
$E_{jk}$ for $(j,k) \neq (1,1)$ or
$Q_{jk}$ for $j \neq 1$.
Thus, by applying \citet[][Theorem 1]{horowitz2000nonparametric}, we obtain the sharp bounds
\begin{align}
\frac{E_{11}Q_{11}p_{11}}{\sum_{k} Q_{1k}p_{1k}+p_{00}+(1-A_{01})p_{01}}
\leq&
\E[Y_2(1)|Y_1(1)=0,G=O]
\notag\\
\leq&
\frac{E_{11}Q_{11}p_{11}+Q_{10}p_{10}+p_{00}+A_{01}p_{01}}{\sum_{k}Q_{1k}p_{1k}+p_{00}+A_{01}p_{01}},
\label{eq:sharp_bounds1}
\end{align}
where
$
A_{01} = \P(Y_2(1)=1|Z_1=0,Z_2=1,G=O).
$
Since $\P(Z_1=0,Z_2=0|G=O) + \P(Z_1=1,Z_2=1|G=O) = \P(W=0|G=O) + \P(W=1|G=O) = 1$, we have $p_{01}=p_{10}=0$.
But then, the sharp bounds \eqref{eq:sharp_bounds1} reduce to
\begin{align*}
\frac{E_{11}Q_{11}p_{11}}{Q_{11}p_{11}+p_{00}}
\leq
\E[Y_2(1)|Y_1(1)=0,G=O]
\leq
\frac{E_{11}Q_{11}p_{11}+p_{00}}{Q_{11}p_{11}+p_{00}}.
\end{align*}
The proof is complete by noting that
$E_{11}=\P(Y_2=1|Y_1=0,W=1,G=O)$,
$Q_{11}=\P(Y_1=0|W=1,G=O)$, and
$p_{jj}=\P(W=j|G=O)$ for each $j \in \{0,1\}$.
%%%%%%%%%%%%%%%%%%%%%%%%%%%%%%%%%%%%%%%
\bigskip\\
\noindent
{Part (ii):}
Let $Z_1$ (respectively, $Z_2$) indicate the even that a researcher observes $Y_1(1)$ (respectively, $Y_2(0)$).
By the law of total probability and then applying the Bayes' theorem, we have
\begin{align*}
&\E[Y_2(0)|Y_1(1)=0,G=O]=
\sum_{j,k} E_{jk} G_{jk}
=
\frac{\sum_{j,k} E_{jk} Q_{jk} p_{jk}}{\sum_{j,k} Q_{jk} p_{jk}}
\end{align*}
where
$
E_{jk} = \P(Y_2(0)=1|Y_1(1)=0,Z_1=j,Z_2=k,G=O),
$
$
G_{jk} = \P(Z_1=j,Z_2=k|Y_1(1)=0,G=O),
$
$
Q_{jk} = \P(Y_1(1)=0|Z_1=j,Z_2=k,G=O),
$
and
$
p_{jk} = \P(Z_1=j,Z_2=k|G=O).
$
Note that we do not observe
$E_{jk}$ for $(j,k) \neq (1,1)$ or
$Q_{jk}$ for $j \neq 1$.
Thus, by applying \citet[][Theorem 1]{horowitz2000nonparametric}, we obtain the sharp bounds
\begin{align}
\frac{E_{11}Q_{11}p_{11}}{\sum_{k} Q_{1k}p_{1k}+p_{00}+(1-A_{01})p_{01}}
\leq&
\E[Y_2(0)|Y_1(1)=0,G=O]
\notag\\
\leq&
\frac{E_{11}Q_{11}p_{11}+Q_{10}p_{10}+p_{00}+A_{01}p_{01}}{\sum_{k}Q_{1k}p_{1k}+p_{00}+A_{01}p_{01}},
\label{eq:sharp_bounds2}
\end{align}
where
$
A_{01} = \P(Y_2(0)=1|Z_1=0,Z_2=1,G=O).
$
Since $\P(Z_1=0,Z_2=1|G=O) + \P(Z_1=1,Z_2=0|G=O) = \P(W=0|G=O) + \P(W=1|G=O) = 1$, we have $p_{00}=p_{11}=0$.
But then, the sharp bounds \eqref{eq:sharp_bounds2} reduce to
\begin{align*}
0
\leq
\E[Y_2(0)|Y_1(1)=0,G=O]
\leq
1
\end{align*}
as claimed.
\end{proof}
%%%%%%%%%%%%%%%%%%%%%%%%%%%%%%%%%%%%%%%

%%%%%%%%%%%%%%%%%%%%%%%%%%%%%%%%%%%%%%%
\subsection{Proof of Corollary \ref{corollary:no_informative}}\label{sec:corollary:no_informative}
%%%%%%%%%%%%%%%%%%%%%%%%%%%%%%%%%%%%%%%
\begin{proof}
The claim follows from Theorem \ref{theorem:non_informative} (ii) and Definition \ref{def:informative}.
\end{proof}
%%%%%%%%%%%%%%%%%%%%%%%%%%%%%%%%%%%%%%%

%%%%%%%%%%%%%%%%%%%%%%%%%%%%%%%%%%%%%%%
\subsection{Proof of Theorem \ref{theorem:data_combination2}}\label{sec:theorem:data_combination2}
%%%%%%%%%%%%%%%%%%%%%%%%%%%%%%%%%%%%%%%
\begin{proof}
Under no restriction on the conditional copula between $Y_2(0)$ and $Y_1(1)$ given $G=O$,
the Fr\'echet-Hoeffding Theorem yields the sharp bounds
\begin{align*}
\max\{\P(Y_2(0)=1|G=O)+\P(Y_1(1)=0|G=O)-1,0\}
\leq
\P(Y_2(0)=1,Y_1(1)=0|G=O)
\\
\leq
\min\{\P(Y_2(0)=1|G=O),\P(Y_1(1)=0|G=O)\}
\end{align*}
for $\P(Y_2(0)=1,Y_1(1)=0|G=O)$.
Therefore, the sharp bounds of
\begin{align*}
\E[Y_2(0)|Y_1(1)=0,G=O]
=&
\frac{\P(Y_2(0)=1,Y_1(1)=0|G=O)}{\P(Y_1(1)=0|G=O)}
\end{align*}
can be written as
\begin{align}
\max\left\{\frac{\E[Y_2(0)|G=O] - \E[Y_1(1)|G=O]}{1-\E[Y_1(1)|G=O]},0\right\}
\leq
\E[Y_2(0)|Y_1(1)=0,G=O]
\notag\\
\leq
\min\left\{\frac{\E[Y_2(0)|G=O]}{1-\E[Y_1(1)|G=O]},1\right\}
\label{eq:sharp_bounds}
\end{align}
The component, $\E[Y_1(1)|G=O]$, of the sharp bounds \eqref{eq:sharp_bounds} can be point-identified by
\begin{align}
\E[Y_1(1)|G=O]
=&
\E[\E[Y_1(1)|X,G=O]|G=O]
\notag\\
=&
\E[\E[Y_1(1)|X,G=E]|G=O]
\notag\\
=&
\E[\E[Y_1|W=1,X,G=E]|G=O]
\label{eq:sharp_bounds:Y11}
\end{align}
where 
the first equality is due to the law of iterated expectations,
the second equality follows from Assumption \ref{ass:EV}, and
the third equality is due to Assumption \ref{ass:IV}.
The other component, $\E[Y_2(0)|G=O]$, of the sharp bounds \eqref{eq:sharp_bounds} is identified by \citet{athey2020combining} as
\begin{align}\label{eq:sharp_bounds:Y20}
\E[Y_2(0)|G=O] = \E[\E[\E[Y_2|Y_1,W=0,X,G=O]|X,G=E]|G=O]
\end{align}
under Assumptions \ref{ass:IV}, \ref{ass:EV}, and \ref{ass:LU}.
Since Assumptions \ref{ass:IV}, \ref{ass:EV}, and \ref{ass:LU} do not impose any restriction on the conditional copula between $Y_2(0)$ and $Y_1(1)$ given $G=O$, the Fr\'echet-Hoeffding bounds \eqref{eq:sharp_bounds} remain sharp under these assumptions.
Therefore, substituting \eqref{eq:sharp_bounds:Y11} and \eqref{eq:sharp_bounds:Y20} in \eqref{eq:sharp_bounds} yields the sharp bounds\small
\begin{align*}
\max\left\{\frac{\E[\E[\E[Y_2|Y_1,W=0,X,G=O]|X,G=E]|G=O] - \E[\E[Y_1|W=1,X,G=E]|G=O]}{1-\E[\E[Y_1|W=1,X,G=E]|G=O]},0\right\}&
\\
\leq
\E[Y_2(0)|Y_1(1)=0,G=O]
\leq
\min\left\{\frac{\E[\E[\E[Y_2|Y_1,W=0,X,G=O]|X,G=E]|G=O]}{1-\E[\E[Y_1|W=1,X,G=E]|G=O]},1\right\}&,
\end{align*}\normalsize
as claimed.
\end{proof}
%%%%%%%%%%%%%%%%%%%%%%%%%%%%%%%%%%%%%%%

%%%%%%%%%%%%%%%%%%%%%%%%%%%%%%%%%%%%%%%
\subsection{Proof of Corollary \ref{corollary:informative}}\label{sec:corollary:informative}
%%%%%%%%%%%%%%%%%%%%%%%%%%%%%%%%%%%%%%%
\begin{proof}
In view of \eqref{eq:sharp_bounds} in the proof of Theorem \ref{theorem:data_combination2} in Appendix \ref{sec:theorem:data_combination2}, we find that the bounds are informative if and only if
$\E[Y_2(0)|G=O] - \E[Y_1(1)|G=O] > 0$ 
or 
$\E[Y_2(0)|G=O] < 1-\E[Y_1(1)|G=O]$.
Therefore, the statement of the corollary follows.
\end{proof}
%%%%%%%%%%%%%%%%%%%%%%%%%%%%%%%%%%%%%%%

%%%%%%%%%%%%%%%%%%%%%%%%%%%%%%%%%%%%%%%
\subsection{Proof of Theorem \ref{theorem:NSD}}\label{sec:theorem:NSD}
%%%%%%%%%%%%%%%%%%%%%%%%%%%%%%%%%%%%%%%
\begin{proof}
First, \citet{athey2020combining} shows
\begin{align*}
\E[Y_2(1)|G=O]-\E[Y_2(0)|G=O]
=
&\E[\E[\E[Y_2|Y_1,W=1,X,G=O]|X,G=E]|G=O]
\\
-
&\E[\E[\E[Y_2|Y_1,W=0,X,G=O]|X,G=E]|G=O].
\end{align*}
under Assumptions \ref{ass:IV}, \ref{ass:EV}, and \ref{ass:LU}.
Second, Assumption \ref{ass:NSD} implies
\begin{align*}
\theta =& \E[Y_2(1)|Y_1(1)=0,G=O] - \E[Y_2(0)|Y_1(1)=0,G=O]\\
=& \E[Y_2(1)|G=O] - \E[Y_2(0)|G=O].
\end{align*}
Combining the above two equations completes a proof of the theorem.
\end{proof}
%%%%%%%%%%%%%%%%%%%%%%%%%%%%%%%%%%%%%%%

%%%%%%%%%%%%%%%%%%%%%%%%%%%%%%%%%%%%%%%
\section{Point Identification under No State Dependence}\label{sec:point_identification}
%%%%%%%%%%%%%%%%%%%%%%%%%%%%%%%%%%%%%%%

The main text focuses on the partial identifiability of the ATETS, $\theta$, under data combination and model restrictions.
This section presents an assumption that allows for the point identification of the ATETS, $\theta$, under the data combination.

Specifically, imposing the following condition of no state dependence, in addition to the three baseline conditions (Assumptions \ref{ass:IV}--\ref{ass:LU}), point-identifies $\theta$.

%%%%%%%%%%%%%%%%%%%%%%%%%%%%%%%%%%%%%%%
\begin{assumption}[No State Dependence] \label{ass:NSD}
    ${}$\\${}$\hspace{1cm}
    $(Y_2(1),Y_2(0)) \indep (Y_1(1),Y_1(0)) | G=O$.
\end{assumption}
%%%%%%%%%%%%%%%%%%%%%%%%%%%%%%%%%%%%%%%

The literature, initiated by Heckman in the early 1980s \citep{heckman1981heterogeneity,heckman1984method}, grapples with interpreting serial correlations in sequential outcomes. Two main theories have been posited: (i) time-varying unobservables (unobserved heterogeneity) influencing sequential outcomes, and (ii) state dependence, where previous unemployment impacts future unemployment chances due to reduced social skill opportunities.

Distinguishing between state dependence and heterogeneity is challenging, yet empirical studies have aimed to clarify this distinction. For example, \citet{JIA2021102004} examined various channels of state dependence in the context of Norwegian tax reforms, finding that state dependence reduces first-year responses to a third of the tax reform's full effect.
\citet{knights2002dynamic} explored Australia's labor market through a random utility model, uncovering evidence of state dependence, indicative of a scarring effect from unemployment.
Important recent work by \cite{torgovitsky2019nonparametric} taking a partial identification approach revealed that nonparametric assumptions attribute at least 30–40 percent of four-month unemployment persistence among high school-educated U.S. men to state dependence.

As mentioned in the main text, positing that there is no state dependence, and that all correlation stems  from heterogeneity in fact point identifies the ATETS, but given the empirical evidence, that might be too heroic an assumption. 
Hence, we present the following point identification result only in this appendix.

%%%%%%%%%%%%%%%%%%%%%%%%%%%%%%%%%%%%%%%
\begin{theorem}[Point Identification under No State Dependence] \label{theorem:NSD}
If Assumptions \ref{ass:IV}, \ref{ass:EV}, \ref{ass:LU}, and \ref{ass:NSD} are satisfied, then $\theta$ is point identified by
\begin{align*}
\theta
=
&\E[\E[\E[Y_2|Y_1,W=1,X,G=O]|X,G=E]|G=O]
\\
-
&\E[\E[\E[Y_2|Y_1,W=0,X,G=O]|X,G=E]|G=O].
\end{align*}
\end{theorem}
%%%%%%%%%%%%%%%%%%%%%%%%%%%%%%%%%%%%%%%

Our proof can be found in Appendix \ref{sec:theorem:NSD}.

%%%%%%%%%%%%%%%%%%%%%%%%%%%%%%%%%%%%%%%
\section{Additional Examples of Imposing Model Selections}\label{sec:additional_model_selections}
%%%%%%%%%%%%%%%%%%%%%%%%%%%%%%%%%%%%%%%

Section \ref{sec:MTR} in the main text focuses on the monotone treatment response (MTR) condition as a leading example of model restrictions on $\mathcal{P}$.
In the current appendix section, we provide three additional examples: stochastic dominance, stationarity, and positively correlated outcomes.

%%%%%%%%%%%%%%%%%%%%%%%%%%%%%%%%%%%%%%%
\subsection{Stochastic Dominance}
%%%%%%%%%%%%%%%%%%%%%%%%%%%%%%%%%%%%%%%

Define the stochastic dominance as follows:
\begin{align}\label{eq:sd}
\P(Y_t(1)=0|X=x,G=g)
\leq
\P(Y_t(0)=0|X=x,G=g)
\end{align}
for each $(x,g) \in \mathcal{X} \times \mathcal{G}$.

We can impose this stochastic dominance condition \eqref{eq:sd} by restricting $\mathcal{P}$ with the requirement that 
\begin{align*}
\sum_{\widetilde y_2(0)=0}^1
\sum_{\widetilde y_1(1)=0}^1 \sum_{\widetilde y_1(0)=0}^1 
\sum_{w = 0}^1
Q_{U|WXG}(\{(0,\widetilde y_2(0),\widetilde y_1(1),\widetilde y_1(0))\}|w,x,g) f_{W|XG}(w|x,g)
\\
\leq
\sum_{\widetilde y_2(1)=0}^1
\sum_{\widetilde y_1(1)=0}^1 \sum_{\widetilde y_1(0)=0}^1 
\sum_{w = 0}^1
Q_{U|WXG}(\{(\widetilde y_2(1),0,\widetilde y_1(1),\widetilde y_1(0))\}|w,x,g) f_{W|XG}(w|x,g)
\end{align*}
and
\begin{align*}
\sum_{\widetilde y_2(1)=0}^1 \sum_{\widetilde y_2(0)=0}^1 
\sum_{\widetilde y_1(0)=0}^1
\sum_{w = 0}^1
Q_{U|WXG}(\{(\widetilde y_2(1),\widetilde y_2(0),0,\widetilde y_1(0))\}|w,x,g) f_{W|XG}(w|x,g)
\\
\leq
\sum_{\widetilde y_2(1)=0}^1 \sum_{\widetilde y_2(0)=0}^1 
\sum_{\widetilde y_1(1)=0}^1
\sum_{w = 0}^1
Q_{U|WXG}(\{(\widetilde y_2(1),\widetilde y_2(0),\widetilde y_1(1),0)\}|w,x,g) f_{W|XG}(w|x,g)
\end{align*}
for all $x \in \mathcal{X}$, $g \in \mathcal{G}$, and $Q_{U|WXG} \in \mathcal{P}$.

%%%%%%%%%%%%%%%%%%%%%%%%%%%%%%%%%%%%%%%
\subsection{Stationarity}
%%%%%%%%%%%%%%%%%%%%%%%%%%%%%%%%%%%%%%%

Define the stationarity as follows:
\begin{align}\label{eq:starionarity}
(Y_2(1),Y_2(0)) \sim (Y_1(1),Y_1(0)) | (W,X,G)=(w,x,g) 
\end{align}
for each $(w,x,g) \in \mathcal{W} \times \mathcal{X} \times \mathcal{G}$.

We can impose this stationarity condition \eqref{eq:starionarity} by restricting $\mathcal{P}$ with the requirement that 
\begin{align*}
\sum_{\widetilde y_1(1)=0}^1 \sum_{\widetilde y_1(0)=0}^1 Q_{U|WXG}(\{(y(1),y(0),\widetilde y_1(1),\widetilde y_1(0))\}|w,x,g)
\\
=
\sum_{\widetilde y_2(1)=0}^1 \sum_{\widetilde y_2(0)=0}^1 Q_{U|WXG}(\{(\widetilde y_2(1),\widetilde y_2(0),y(1),y(0))\}|w,x,g)
\end{align*}
for all $(y(1),y(0)) \in \{0,1\}^2$, $w \in \mathcal{W}$, $x \in \mathcal{X}$, $g \in \mathcal{G}$, and $Q_{U|WXG} \in \mathcal{P}$.

%%%%%%%%%%%%%%%%%%%%%%%%%%%%%%%%%%%%%%%
\subsection{Positively Correlated Outcomes}
%%%%%%%%%%%%%%%%%%%%%%%%%%%%%%%%%%%%%%%

Adapting \citet{vikstrom2018bounds} to our framework, define the positively correlated outcomes as follows:
\begin{align}
&\P(Y_2(0)=1|Y_1(1)=0,Y_1(0)=0,X=x,G=g)
\notag\\
&\leq
\P(Y_2(0)=1|Y_1(1)=1,Y_1(0)=0,X=x,G=g)
\label{eq:pco}
\end{align}
for each $(x,g) \in \mathcal{X} \times \mathcal{G}$.

We can impose this positively correlated outcome condition \eqref{eq:pco} by restricting $\mathcal{P}$ with the requirement that
\begin{align*}
&\frac{\sum_{\widetilde y_2(1)=0}^1 \sum_{w=0}^1 Q_{U|WXG}(\{(\widetilde y_2(1),1,0,0)\}|w,x,g) f_{W|XG}(w|x,g)}{\sum_{\widetilde y_2(1)=0}^1 \sum_{\widetilde y_2(0)=0}^1 \sum_{w=0}^1 Q_{U|WXG}(\{(\widetilde y_2(1),\widetilde y_2(0),0,0)\}|w,x,g) f_{W|XG}(w|x,g)}
\\
&\leq
\frac{\sum_{\widetilde y_2(1)=0}^1 \sum_{w=0}^1 Q_{U|WXG}(\{(\widetilde y_2(1),1,1,0)\}|w,x,g) f_{W|XG}(w|x,g)}{\sum_{\widetilde y_2(1)=0}^1 \sum_{\widetilde y_2(0)=0}^1 \sum_{w=0}^1 Q_{U|WXG}(\{(\widetilde y_2(1),\widetilde y_2(0),1,0)\}|w,x,g) f_{W|XG}(w|x,g)}
\end{align*}
for all $x \in \mathcal{X}$, $g \in \mathcal{G}$, and $Q_{U|WXG} \in \mathcal{P}$,
where $f_{W|XG}$ denotes the conditional probability mass function.

%%%%%%%%%%%%%%%%%%%%%%%%%%%%%%%%%%%%%%%
\section{Computation of the Sharp Identified Set}\label{sec:computation}
%%%%%%%%%%%%%%%%%%%%%%%%%%%%%%%%%%%%%%%

A typical element $Q_{U|WXG}$ of $\mathcal{P}$ can be represented by the finite-dimensional parameter vector $q: \mathcal{U} \times \mathcal{W} \times \mathcal{X} \times \mathcal{G} \rightarrow [0,1]$, where $q(\cdot,\cdot,\cdot,\cdot|w,x,g)$ resides in the 15-dimensional simplex $\Delta$ for each $(w,x,g) \in \mathcal{W} \times \mathcal{X} \times \mathcal{G}$.\footnote{The dimension of the simplex is 15 because $|\mathcal{U}|=16$.}
Let $\overline\Delta = \times_{(w,x,g) \in \mathcal{W} \times \mathcal{X} \times \mathcal{G}} \Delta$.
With this representation, Theorem \ref{theorem:general} imposes the restrictions
\begin{align}
\P(Y_1=l|W=0,X=x,G=E) \leq& \sum_{i,j,k} q(i,j,k,l|0,x,E) \text{ for all } l, \label{eq:constraint_artstein1}\\
\P(Y_1=k|W=1,X=x,G=E) \leq& \sum_{i,j,l} q(i,j,k,l|1,x,E) \text{ for all } k, \label{eq:constraint_artstein2}\\
\P(Y_2=j, Y_1=l|W=0,X=x,G=O) \leq& \sum_{i,k} q(i,j,k,l|0,x,O)  \text{ for all } j,l, \text{ and}\label{eq:constraint_artstein3}\\
\P(Y_2=i, Y_1=k|W=1,X=x,G=O) \leq& \sum_{j,l} q(i,j,k,l|1,x,O)  \text{ for all } i,k \label{eq:constraint_artstein4}
\end{align}
for all $x \in \mathcal{X}$.

Corollary \ref{corollary:identified_set} thus implies that the upper and lower bounds of the sharp identified set of the ATETS, $\theta$, can be obtained by the maximum and minimum values, respectively, of
\begin{align*}
\Phi(q) = \frac{\sum_{i,j,l,w,x} (i-j) q(i,j,0,l)|w,x,O)) f_{WX|G}(w,x|O)}{\sum_{i,j,l,w,x} q(i,j,0,l|w,x,O))f_{WX|G}(w,x|O)},
\end{align*}
subject to the constraints \eqref{eq:constraint_artstein1}--\eqref{eq:constraint_artstein4}.

Furthermore, by \eqref{eq:IV}, the experimental internal validity (Assumption \ref{ass:IV}) can be imposed with the additional constraints
\begin{align}\label{eq:constraint_IV}
q(i,j,k,l|0,x,E) = q(i,j,k,l|1,x,E).
\end{align}
for all $(i,j,k,l) \in \mathcal{U}$ and $x \in \mathcal{X}$.
Similarly, by \eqref{eq:EV}, the external validity (Assumption \ref{ass:EV}) can be imposed with the additional constraints
\begin{align}\label{eq:constraint_EV}
\sum_{w=0}^1 q(i,j,k,l|w,x,E)f_{W|XG}(w|x,E) = \sum_{w=0}^1 q(i,j,k,l|w,x,O) f_{W|XG}(w|x,O).
\end{align}
for all $(i,j,k,l) \in \mathcal{U}$ and $x \in \mathcal{X}$, where $f_{W|XG}$ denotes the conditional probability mass function.

By \eqref{eq:lu1}--\eqref{eq:lu2}, the latent unconfoundedness (LU) can be imposed by
\begin{align}
&\sum_{\tilde i}\sum_{\tilde k} q(\tilde i,j,\tilde k,l|1,x,O) \cdot \sum_{\tilde i}\sum_{\tilde j}\sum_{\tilde k} q(\tilde i,\tilde j,\tilde k,l|0,x,O)
\notag\\
=&
\sum_{\tilde i}\sum_{\tilde k} q(\tilde i,j,\tilde k,l|0,x,O) \cdot \sum_{\tilde i}\sum_{\tilde j}\sum_{\tilde k} q(\tilde i,\tilde j,\tilde k,l|1,x,O)
\text{ for all } j, l, \text{ and }
\label{eq:constraint_LU1}\\
&\sum_{\tilde j}\sum_{\tilde l} q(i,\tilde j,k,\tilde l|1,x,O) \cdot \sum_{\tilde i}\sum_{\tilde j}\sum_{\tilde l} q(\tilde i,\tilde j,k,\tilde l|0,x,O)
\notag\\
=&
\sum_{\tilde j}\sum_{\tilde l} q(i,\tilde j,k,\tilde l|0,x,O) \cdot \sum_{\tilde i}\sum_{\tilde j}\sum_{\tilde l} q(\tilde i,\tilde j,k,\tilde l|1,x,O) \text{ for all } i,k
\label{eq:constraint_LU2}
\end{align}
for all $x \in \mathcal{X}$.

By \eqref{eq:MTR1}--\eqref{eq:MTR2}, the monotone treatment response (MTR) can be imposed by 
\begin{align}
q(i,j,0,1|x,w,g)=0 & \text{ for all } i,j, \text{ and } \label{eq:constraint_MTR1}\\
q(0,1,k,l|x,w,g)=0 & \text{ for all } k,l \label{eq:constraint_MTR2}
\end{align}
for all $(w,x,g) \in \mathcal{W} \times \mathcal{X} \times \mathcal{G}$. 

In summary, the upper bound of the sharp identified set is obtained by the constrained optimization problem:
\begin{align*}
\max_{q \in \overline\Delta} & \ \Phi(q) \\
\text{s.t.} & \text{ \eqref{eq:constraint_artstein1}--\eqref{eq:constraint_artstein4}}\\
\text{(s.t.} &\text{ \eqref{eq:constraint_IV} if the internal validity (IV) is imposed)}\\
\text{(s.t.} &\text{ \eqref{eq:constraint_EV} if the external validity (EV) is imposed)}\\
\text{(s.t.} &\text{ \eqref{eq:constraint_LU1}--\eqref{eq:constraint_LU2} if the latent unconfoundedness (LU) is imposed)}\\
\text{(s.t.} &\text{ \eqref{eq:constraint_MTR1}--\eqref{eq:constraint_MTR2} if the monotone treatment response (MTR) is imposed)}
\end{align*}

%%%%%%%%%%%%%%%%%%%%%%%%%%%%%%%%%%%%%%%
%\section{Data}\label{sec:data}
%%%%%%%%%%%%%%%%%%%%%%%%%%%%%%%%%%%%%%%
\section{Data}\label{sec:data_job_training}
%%%%%%%%%%%%%%%%%%%%%%%%%%%%%%%%%%%%%%%

This section presents details of the data sets we use in the empirical application in Section \ref{sec:application_job_training}.
Following \citet{athey2019surrogate}, both experimental and observational data are constructed from the California GAIN program, a randomized control trial (RCT) designed to evaluate the impact of job training programs on long-term labor market outcomes. 
Specifically, data from Riverside serve as the experimental sample, emphasizing a ``jobs first'' approach. The observational sample combines data from three other locations (Alameda, Los Angeles, and San Diego).

The Greater Avenues to Independence (GAIN) program was a job assistance initiative launched in California during the 1980s, aimed at enabling welfare recipients to secure employment. Its evaluation, conducted by MDRC, involved a randomized trial across six counties, with a particular focus on the Riverside trial due to its notable impact on earnings. Unlike other sites that prioritized human capital development through training, Riverside adopted a ``jobs first'' strategy, urging participants to accept any available job to expedite reentry into the labor force.

The study explores long-term labor market outcomes (employment, earnings, and aid receipt) over thirty-six quarters post-random-assignment, primarily utilizing data from Riverside as the experimental sample and data from Alameda, Los Angeles, and San Diego as the observational sample. This approach allowed for a comparison of outcomes based on initial employment, earnings, and aid, as well as pre-treatment variables such as individual characteristics and previous labor market experiences.

In Riverside, out of 5,445 participants, 4,405 were in the treatment group, and 1,040 were in the control group, not receiving additional services. The analysis, leveraging data from \citet{hotz2005predicting}, which tracked participants for nine years, showed that while the Riverside GAIN program initially boosted employment rates and earnings significantly, these effects diminished over time.

%{\color{red} TO DO}

%%%%%%%%%%%%%%%%%%%%%%%%%%%%%%%%%%%%%%%
\setlength{\baselineskip}{6.8mm}
\bibliographystyle{apalike}
\bibliography{reference}
%%%%%%%%%%%%%%%%%%%%%%%%%%%%%%%%%%%%%%%
\end{document}